\documentclass[prb,twocolumn,showpacs,amsmath,amssymb]{revtex4}
\usepackage{graphicx}
\usepackage{tabularx}

\begin{document}

\title{Steady State of Counterflow Quantum Turbulence: Vortex filament Simulation with the Full Biot--Savart Law}


\author{Hiroyuki Adachi}
\affiliation{Department of Physics, Osaka City University, Sumiyoshi-Ku, Osaka 558-8585, Japan}
\author{Shoji Fujiyama}
\affiliation{Department of Physics, Osaka City University, Sumiyoshi-Ku, Osaka 558-8585, Japan}
\author{Makoto Tsubota}
\affiliation{Department of Physics, Osaka City University, Sumiyoshi-Ku, Osaka 558-8585, Japan}

\date{\today}

\begin{abstract}
We perform a numerical simulation of quantum turbulence produced by thermal counterflow in superfluid $^4$He by using the vortex filament model with the full Biot--Savart law.
The pioneering work of Schwarz has two shortcomings:
it neglects the non-local terms of the Biot--Savart integral (known as the localized induction approximation, LIA) and it employs an unphysical mixing procedure to sustain the statistically steady state of turbulence.
For the first time we have succeeded in generating the statistically steady state under periodic boundary conditions without using the LIA or the mixing procedure.
This state exhibits the characteristic relation $L=\gamma^2 v_{ns}^2$ between the line-length density $L$ and the counterflow relative velocity $v_{ns}$ and there is quantitative agreement between the coefficient $\gamma$ and some measured values. The parameter $\gamma$ and some anisotropy parameters are calculated as functions of temperature and the counterflow relative velocity. The numerical results obtained using the full Biot--Savart law are compared with those obtained using the LIA. The LIA calculation constructs a layered structure of vortices and does not proceed to a turbulent state but rather to another anisotropic vortex state; thus, the LIA is not suitable for simulations of turbulence.
\end{abstract}


\maketitle

\section{\label{sec:intro}introduction}
Quantum turbulence \cite{tsubota,tsubota_jpsj}, which is the disordered motion of a tangle of quantized vortices, has been investigated since pioneering thermal counterflow experiments by Vinen in the late 1950s \cite{vinen}. Many experimental, theoretical, and numerical studies have advanced our understanding of counterflow turbulence.
 Many researchers have recently focused on quantum turbulence near 0\ K. One of their main motivations is that quantum turbulence at 0\ K exhibits some similarities with turbulence in ordinary fluids \cite{tsubota}.
 A typical example is Kolmogorov's law, which is the most important statistical law of classical turbulence and which has been numerically confirmed in quantum turbulence too \cite{kolmogorov}.
 In contrast, turbulence at finite temperatures (e.g., counterflow turbulence) is a form of motion peculiar to two-fluid hydrodynamics \cite{landau} and has no direct analog with turbulence in an ordinary viscous fluid.
 For this reason, turbulence at finite temperatures is not currently investigated as much as turbulence at 0 K.
However, the physics of finite-temperature turbulence is far from fully understood \cite{tough}.
Recently, particle image velocimetry (PIV) experiments \cite{bewley,paoletti,zhang}, which is a technique for visualizing fluid flow using small particles, have been performed to investigate counterflow of superfluid $^4$He. In order to interpret these experiments, we need to understand the microscopic vortex dynamics in thermal counterflow.
Our current understanding of counterflow turbulence is still deficient. Hence, in this introduction we briefly review research of counterflow turbulence and reveal some important unresolved problems.

 According to the two-fluid model, superfluid $^4$He consists of an intimate mixture of two fluid components: a viscous normal fluid and an inviscid superfluid.
 The density and velocity of the normal fluid are respectively denoted by $\rho_n$ and ${\bf v}_n$, while the superfluid density and velocity are respectively denoted by $\rho_s$ and ${\bf v}_s$.
The total density, $\rho = \rho_s+\rho_n$, is approximately temperature independent, but the relative proportions of the normal fluid and the superfluid, $\rho_s/\rho$ and $\rho_n/\rho$, depends strongly on the temperature.
In this system, any rotational motion of a superfluid is sustained only by quantized vortices, which have the quantum circulation $\kappa=h/m_4$, where $h$ is Planck's constant and $m_4$ is the mass of a $^4$He atom.

 A thermal counterflow, which is internal convection produced by a temperature gradient, is explained by this two-fluid model.
 In a counterflow, entropy and heat are carried only by the normal fluid component. Hence, if a heat current is applied to the closed end of channel, then the normal fluid will flow from warmer areas to cooler areas while the superfluid will flow in the opposite direction to conserve the total mass. In this way, counterflow is induced with a relative velocity between the superfluid and the normal fluid of ${\bf v}_{ns}={\bf v}_{n}-{\bf v}_{s}$.
  However, superflow becomes dissipative (superfluid turbulence) above a certain critical counterflow velocity.
 The concept of superfluid turbulence was introduced by Feynman \cite{feynman} who stated that the turbulent state consists of a disordered set of quantized vortices, called a vortex tangle.

 This idea was further developed by Vinen.
 In order to describe amplification of a temperature difference at the ends of a capillary retaining thermal counterflow, Gorter and Mellink introduced some additional interactions between the normal fluid and superfluid (mutual friction) \cite{gorter}.
 Through experimental studies of the second-sound attenuation, Vinen considered this Gorter--Mellink mutual friction in relation to the macroscopic dynamics of the vortex tangle \cite{vinen}. Assuming homogeneous superfluid turbulence (actually, counterflow turbulence is anisotropic), Vinen obtained an equation for the evolution of the vortex line density (VLD) $L(t)$, which we call Vinen's equation:
\begin{equation}
\frac{dL}{dt}=\alpha|{\bf v}_{ns}|L^{3/2}-\chi_2 \frac{\kappa}{2\pi}L^2,
\end{equation}
where $\alpha$ and $\chi_2$ are temperature-dependent parameters. The first term represents the energy injection from the normal fluid to the vortices. The second term denotes the energy dissipation of vortices due to reconnection between vortices.
 The first and second terms indicate the growth and the degeneration of a vortex tangle, respectively.
 Therefore, after the growth period of the VLD, the vortex tangle enters a statistically steady state.
 In the steady state, the VLD is obtained by setting $dL/dt$ equal to zero, which gives
\begin{equation}
L=\gamma^2v_{ns}^2,
\end{equation}
where $\gamma$ is a temperature-dependent parameter.
This relation is able to describe well a large number of observations of stationary cases \cite{tough}.

 On the other hand, the nonlinear and nonlocal dynamics of vortices have long delayed progress in achieving a microscopic understanding of quantum turbulence.
 It was Schwarz who made a breakthrough \cite{schwarz85,schwarz88}.
 He investigated counterflow turbulence using the vortex filament model and dynamic scaling \cite{schwarz88}. 
The observable quantities obtained by his calculation agree well with the experimental results for the steady state of vortex tangles. This study confirmed the idea proposed by Feynman that superfluid turbulence consists of a quantized vortex tangle. However, thermal counterflow turbulence is far from being perfectly understood.
 The numerical simulation of Schwarz has serious defects.
 One is that calculations are performed under the localized induction approximation (LIA), which neglects interactions between vortices. Schwarz reported that as a result the layer structure is constructed gradually when periodic boundary conditions are applied. Of course, this behavior differs from experimental observations.
 In order to remedy this, an unphysical, artificial mixing procedure was employed, in which half the vortices are randomly selected to be rotated by 90$^\circ$ around the axis defined by the flow velocity. It is only this method that enables the steady state to be sustained under periodic boundary conditions.
 These defects cause us to conjecture that the LIA is unsuitable due the absence of interactions between vortices.
To understand counterflow turbulence properly, simulations have to be performed using the full Biot--Savart law without using the artificial mixing procedure.

 The contents of this paper are as follows. Section I\hspace{-.1em}I describes the equations of motion of vortices and the numerical calculation method. In Sec. I\hspace{-.1em}I\hspace{-.1em}I, we show numerical simulations of counterflow turbulence by the full Biot--Savart law and some physical parameters such as the VLD and anisotropy parameters. Section I\hspace{-.1em}V compares the results obtained using the full Biot--Savart law with those obtained using the LIA, and shows that the LIA is unsuitable. Section V is devoted to conclusions and discussions.
  
\section{EQUATIONS OF MOTION}
 A quantized vortex is represented by a filament passing through a fluid and has a definite direction corresponding to its vorticity. Except for the thin core region, a superflow velocity field has a classically well-defined meaning and can be described by ideal fluid dynamics. The velocity produced at a point ${\bf r}$ by a filament is given by the Biot--Savart expression:
\begin{equation}
{\bf v}_{\omega}=\frac{\kappa}{4\pi}\displaystyle \int_{\cal L} \frac{({\bf s}_1-{\bf r})\times d{\bf s}_1}{|{\bf s}_1-{\bf r}|^3}.
\end{equation}
 The filament is represented in parametric form ${\bf s}={\bf s}(\zeta,t)$, where ${\bf s}_1$ refers to a point on the filament and the integration is performed along the filament. Helmholtz's theorem for a perfect fluid states that the vortex moves with the superfluid velocity at the point. Attempting to calculate the velocity ${\bf v}_{\omega}$ at a point ${\bf r}={\bf s}$ on the filament causes the integral diverge as ${\bf s}_1\rightarrow {\bf s}$. To avoid this, we divide the velocity $\dot{{\bf s}}$ of the filament at the point {\bf s} into two components: \cite {schwarz85}
\begin{equation}
{\dot {\bf s}}=\frac{\kappa}{4\pi}{\bf s}'\times{\bf s}'' \ln \biggl( \frac{2(l_+l_-)^{1/2}}{e^{1/4}a_0} \biggr)+\frac{\kappa}{4\pi}\displaystyle\int^{'}_{\cal L}\frac{({\bf s}_1-{\bf s})\times d{\bf s}_1}{|{\bf s}_1-{\bf s|^3}},
\end{equation}
where the prime denotes derivatives with respect to the arc length $\zeta$, $a_0$ is a cutoff parameter corresponding to the vortex core radius, and $l_+$ and $l_-$ are the lengths of the two adjacent line elements connected to point ${\bf s}$. The first term denotes the localized induction field arising from a curved line element acting on itself. The second term represents the non-local field obtained by performing the Biot--Savart integral along the rest of the filament. 

The LIA, which has been used in several studies \cite{schwarz85,schwarz88,araki00, kondaurova}, involves neglecting the second non-local term in Eq. (4). The equation for the LIA is often written as
\begin{equation}
{\dot{\bf s}}=\beta{\bf s}'\times {\bf s}''.
\end{equation}
Here the coefficient $\beta$ is defined by 
\begin{equation}
\beta = \frac{\kappa}{4\pi}\ln \biggl( \frac{c\langle R \rangle}{a_0} \biggr),
\end{equation}
 where $c$ is a constant of order unity and $(l_+ l_-)^{1/2}$ is replaced by the characteristic radius $\langle R \rangle $ of curvature of the vortex lines.
In contrast, calculations without the LIA are referred to as full Biot--Savart calculations. 

 When counterflow is applied, the applied superfluid velocity ${\bf v}_s$ is added to ${\bf v}_{\omega}$, and the total velocity ${\dot{\bf s}}_0$ of the vortex filament without dissipation is:
\begin{equation}
\begin{split}
{\dot {\bf s}_0}=&\frac{\kappa}{4\pi}{\bf s}'\times{\bf s}'' \ln \biggl( \frac{2(l_+l_-)^{1/2}}{e^{1/4}a_0} \biggr)+\frac{\kappa}{4\pi}\displaystyle\int^{'}_{\cal L}\frac{({\bf s}_1-{\bf s})\times d{\bf s}_1}{|{\bf s}_1-{\bf s|^3}} \\ &+{\bf v}_s.
\end{split}
\label{s0}
\end{equation}
At finite temperatures the mutual friction due to the interaction between the vortex core and the normal-fluid flow ${\bf v}_n$ is taken into account. The velocity of a point {\bf s} is then given by \cite{schwarz85}
\begin{equation}
{\dot{\bf s}}={\dot{\bf s}_0}+\alpha {\bf s}'\times ({\bf v}_n-{\dot{\bf s}}_0)-\alpha '{\bf s}'\times[{\bf s}'\times({\bf v}_n-{\dot{\bf s}}_0)],
\label{s}
\end{equation}
where $\alpha$ and $\alpha'$ are the temperature-dependent coefficients, and ${\dot{\bf s}}_0$ is calculated from Eq. (\ref{s0}). Since $\alpha'$ is small compared with $\alpha$, some authors\cite{schwarz85,schwarz88,araki00} neglect $\alpha'$.
In this study, we take into account both $\alpha'$ and $\alpha$. Table \ref{friction} presents the mutual fiction parameters used in this study \cite{schwarz85}.
\begin{table}
\begin{center}

\begin{tabular}{ccc}
\hline
    T (K)   & $\alpha$    & $\alpha'$  \\ \hline
        1.3 & 0.036       & 0.014 \\
        1.6 & 0.098       & 0.016 \\
        1.9 & 0.21        & 0.009 \\
        2.1 & 0.50        & -0.03 \\
\hline
\end{tabular}
\caption{Friction coefficients. \label{friction}}
\end{center}
\end{table}

 The Gross--Pitaevskii (GP) model might be considered more appropriate than the vortex filament model. In contrast with the vortex filament model (which is used by several groups including us)\cite{schwarz85,schwarz88,araki00,kondaurova}, the GP model can represent phenomena associated with the vortex core including reconnection, nucleation, and annihilation. However, no methods have been established for treating mutual friction in the GP model, and huge calculations are necessary to obtain statistics such as the VLD. In contrast, the vortex filament model can incorporate the effect of mutual friction phenomenologically using the experimentally observable parameters $\alpha$ and $\alpha'$, and it has a lower computational cost for dense vortices. For these reasons, the vortex filament model is more suitable for numerically simulating counterflow turbulence.

 Needless to say, mutual friction plays an important role in counterflow turbulence. Let us assume the LIA with Eq. (5) and neglect the term with $\alpha'$. Then, Eqs. (7) and (8) are reduced to 
\begin{equation}
{\dot {\bf s}}=\beta {\bf s}'\times {\bf s}'' +{\bf v}_s + \alpha {\bf s}' \times ({\bf v}_n-{\bf v}_s -\beta{\bf s}' \times {\bf s}'').
\end{equation}
If mutual friction is absent, the dynamics due to only the self-induced velocity conserves the total line length of vortices. When mutual friction is present and the counterflow ${\bf v}_{ns} $ flows against the local self-induced velocity $\beta {\bf s}'\times{\bf s}''$, the mutual friction always shrinks the vortex line locally. On the other hand, the relative flow along the self-induced velocity yields a critical radius of curvature given by
\begin{equation}
R_c \simeq \frac{\beta}{v_{ns}}.
\end{equation}
 When the local radius $R$ at a point on a vortex is smaller than $R_c$, the vortex will shrink locally, while the vortex will balloon out when $R>R_c$. Thus, mutual friction causes the vortex line length to both grow and decay. This dual role of mutual friction sustains the steady state of counterflow turbulence, ensuring that a highly curved structure whose local radius of curvature is less than $R_c$ will be smoothed out. 

 Some important quantities that are useful for characterizing the vortex tangle are introduced below \cite{schwarz88}. The VLD is 
\begin{equation}
L=\frac{1}{\Omega}\int_{\cal L}d\xi,
\end{equation}
where the integral is performed at all vortices in the sample volume $\Omega$. The anisotropy of the vortex tangle that is formed under the counterflow ${\bf v}_{ns}$ is represented by the dimensionless parameters
\begin {equation}
I_{\|}=\frac{1}{\Omega L} \displaystyle\int^{}_{\cal L}[1-({\bf s}'\cdot{\hat {\bf r}_{\|}})^2]d\xi,
\end {equation}
\begin {equation}
I_{\bot}=\frac{1}{\Omega L} \displaystyle\int^{}_{\cal L}[1-({\bf s}'\cdot{\hat {\bf r}_{\bot}})^2]d\xi,
\end {equation}
\begin{equation}
I_l{\hat {\bf r}_{\|}}=\frac{1}{\Omega L^{3/2}}\int^{}_{\cal L} {\bf s}'\times {\bf s}''d \xi.
\end{equation}
 Here, ${\hat {\bf r}}_{\|}$ and ${\hat {\bf r}}_{\bot}$ represent unit vectors parallel and perpendicular to the ${\bf v}_{ns}$ direction, respectively. Symmetry generally yields the relation $I_{\|}/2+I_{\bot}=1$. If the vortex tangle is isotropic, the averages of these parameters are ${\bar I_{\|}}={\bar I_{\bot}}=2/3$ and ${\bar I}_l=0$. At the other extreme, if the tangle consists entirely of curves lying in planes normal to ${\bf v}_{ns}$, ${\bar I_{\|}}=1$ and ${\bar I_{\bot}}=1/2$.

 A numerical study of an incompressible Navier--Stokes fluid revealed that the close interaction of two vortices leads to their reconnection, chiefly because of the viscous diffusion of the vorticity \cite{boratav}. Koplik and Levine directly solved the GP equation to show that two close quantized vortices reconnect even in a superfluid \cite{koplik}. Our numerical method for vortex filaments cannot represent the reconnection process itself. Hence, we reconnect vortices that pass within the space resolution $\Delta \xi$ with unit probability. Every vortex initially consists of a string of points at regular intervals of $\Delta \xi$. When a point on a vortex approaches another point on another vortex more closely than the fixed space resolution $\Delta \xi$, we join these two points and reconnect the vortices. This reconnection procedure is standard in the vortex filament model, but a different procedure is used in some studies \cite{kondaurova}. We discuss this in Sec. V.

 In this study, all calculations are performed under the following conditions.
The numerical space resolution is $\Delta \xi=8.0\times 10^{-4}\,{\rm cm}$, and the time resolution is $\Delta t=1.0\times 10^{-4}\,{\rm s}$
 To integrate the equation of motion given by Eq. (\ref{s}) with respect to time we used the fourth-order Runge--Kutta method. The computing box is $0.1\times 0.1 \times 0.1\, {\rm cm^3}$. We usually start with an initial vortex configuration of six vortex rings, as shown in Fig. \ref{t19}(a).
If the vortex ring is smaller than $R_c$ in Eq. (10), it always shrinks, finally disappearing. Hence in our simulation, we make vortices that are shorter than $\Delta l =7\times\Delta \xi=5.6\times10^{-3}\, {\rm cm}$ to vanish. This cut-off line length is determined to satisfy $\Delta l<R_c$, since a vortex longer than $R_c$ has the possibility of expanding and causing the VLD to increase. 

\section{NUMERICAL SIMULATION OF COUNTERFLOW TURBULENCE}
\begin{figure}[h]
  \begin{center}
 \scalebox{0.5}{\includegraphics{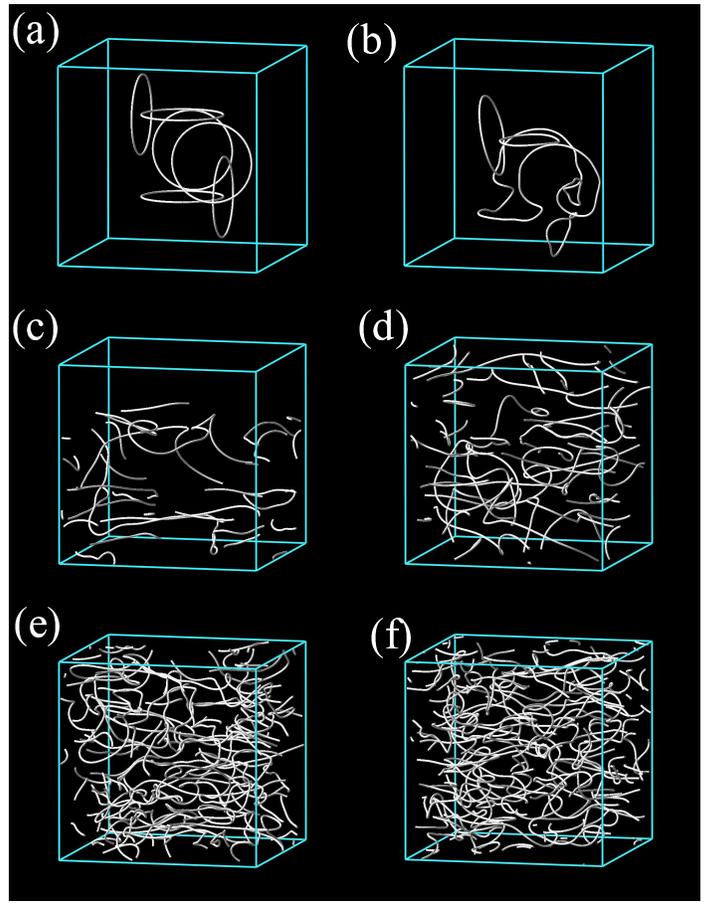}}
 \caption{(Color online). Development of a vortex tangle by the full Biot--Savart calculation in a periodic box with a size of 0.1 cm. Here, the temperature is $T=1.9\,{\rm K}$ and the counterflow velocity $v_{ns}=0.572\,{\rm cm/s}$ is along the vertical axis. (a) $t=0\,{\rm s}$, (b) $t=0.05\,{\rm s}$, (c) $t=0.5\,{\rm s}$, (d) $t=1.0\,{\rm s}$, (e) $t=3.0\,{\rm s}$, (f) $t=4.0\,{\rm s}$.\label{t19}} 
  \end{center} 
 \end{figure}

\begin{figure}[h]
\begin{center}
\includegraphics[width=0.45 \textwidth]{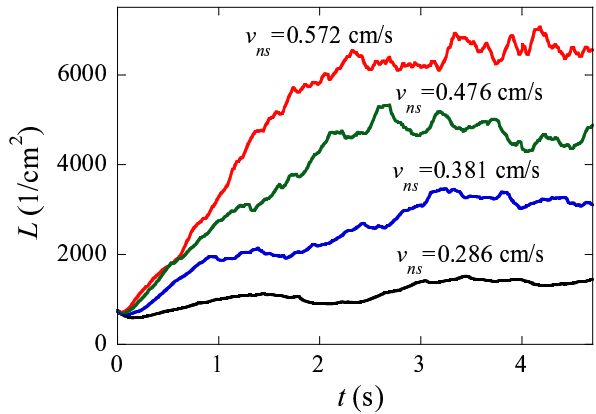}
\caption{Vortex line density as a function of time for four different counterflow velocities. \label{T19_line_t}} 
\end{center} 
\end{figure}

\begin{figure}[h]
\begin{center}
\scalebox{0.5}{\includegraphics{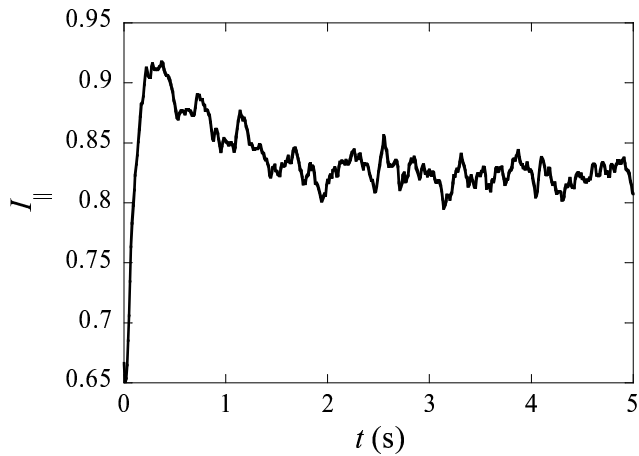}}
\caption{Anisotropy parameter as a function of time. Here, $T=1.9\,{\rm K}$ and counterflow velocity $v_{ns}=0.572\,{\rm cm/s}$.\label{ani_t}} 
\end{center}
\end{figure}

In this section, we present numerical simulations of counterflow turbulence using the full Biot--Savart law under periodic boundary conditions.
 Figure \ref{t19} shows a typical result with the time evolution of the VLD shown in Fig. \ref{T19_line_t} and the anisotropy parameter shown in Fig. \ref{ani_t}. 
The initial configuration consists of six vortex rings. 

In the first stage ($0 \leq t \leq 0.4\, {\rm s}$), the critical radius $R_c$ in Eq. (10) determines the vortex destiny. Vortex ring sections in which the radius of curvature exceeds $R_c$, expand in the direction perpendicular to ${\bf v}_{ns}$ through mutual friction, while small vortex rings shrink. Thus, vortices evolve and become anisotropic, as shown in Figs. \ref{T19_line_t} and \ref{ani_t}. At the end of this stage, large vortices appear that are comparable to the system size under periodic boundary conditions.
These vortices survive with a large radius of curvature, and continuously generate small vortices by reconnections in the subsequent stages so that they function as "vortex mills"\cite{schwarz93}.

In the second stage ($0.4 < t \leq 2.0\, {\rm s}$), vortex tangles undergo continuous evolution despite the decreasing anisotropy.
As vortex rings expand, reconnections between vortices occur frequently. Reconnections generate vortices with various curvatures, resulting in them shrinking and expanding as discussed in the first stage. Local sections with a small radius of curvature formed by reconnections have an almost isotropic self-induced velocity, which prevents the vortices from lying perpendicular to ${\bf v}_{ns}$, and reduces $I_{\|}$. In addition, as the VLD increases, vortex expansion becomes slower than in the first stage because the reconnection distorts vortices, which prevents a vortex from smoothly expanding.

In the third stage ($t > 2.0\,{\rm s}$), the statistically steady state is realized by the competition between the growth and decay of a vortex tangle.
The growth mechanism is still vortex expansion through mutual friction.
The decay mechanism either creates vortices with local radii of curvature smaller than $R_c$ or the self-induced velocity is oriented in the opposite direction to ${\bf v}_{ns}$ after reconnections, as discussed in Sec. I\hspace{-.1em}I. The increasing VLD causes more reconnections so that the decay mechanism becomes effective. When the VLD has increased sufficiently, the two mechanisms begin to compete so that the vortex tangle enters the statistically steady state. The LIA calculation cannot realize this competition, as discussed in Sec. I\hspace{-.1em}V, which shows that vortex interaction is essential for creating a steady state.


\begin{figure}[h]
\begin{center}
\scalebox{0.44}{\includegraphics{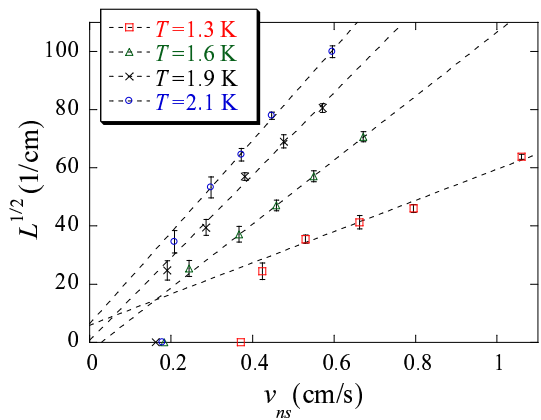}}
\caption{The steady state vortex line density $L(t)$ as a function of the counterflow velocity $v_{ns}$. The error bars represent the standard deviation. \label{vinen}} 

\end{center}
\end{figure} 

 The steady state is known to exhibit the characteristic relation $L=\gamma^2 v_{ns}^2$ as discussed in Sec. I. Our steady states almost satisfy this relation when $v_{ns}$ and $L$ are relatively large, as shown in Fig. \ref{vinen}. In Table \ref{gamma} we show the parameter $\gamma$ as a function of $T$. Our results quantitatively agree with the experimental observations of Childers and Tough \cite {tough,childers}. Additionally, there is a critical velocity of turbulence, below which vortices disappear. This critical velocity has been measured in many previous studies \cite{tough,childers74,deHass}; it is given by 
\begin{equation}
v_{ns,c}\approx \frac{2.5+1.44\sigma}{\gamma d},
\end {equation}
where $d$ is the channel size of the experimental system and $\sigma$ is a constant of order unity. In our simulation, the system size may be taken to be the size of the periodic box. Then, Eq. (15) gives $v_{ns,c} \sim 0.1\, {\rm cm}$, which is almost consistent with our numerical results. However, the temperature dependence of $ v_{ns,c}$ should be discussed. Equation (15) states that $v_{ns,c}$ should decrease with $T$, which differs from the behavior in Fig. \ref{vinen}. Our numerical results show that $v_{ns,c}$ decreases with $T$ below $1.9\, {\rm K}$ but increases at $2.1\, {\rm K}$ slightly. This is because the strong mutual friction makes the vortices so anisotropic that they cannot form enough reconnections with other vortices, and so become degenerate.

\begin{center}
\begin{table}
\begin{tabular}{cccc}
\hline
$T$ (K)& $\gamma_{num}({\rm s/cm^2})$ & $\gamma_{exp}({\rm s/cm^2})$ & $I_{\|}$ \\ \hline
        1.3 & 53.5    &   59  &    0.738 \\
        1.6 & 109.6   &   93  &    0.771 \\
        1.9 & 140.1   &   133 &    0.820 \\
        2.1 & 157.3   &   - &    0.901 \\
\hline
\end{tabular}
\caption{Line density coefficients $\gamma$ and the anisotropy parameter $I_{\|}$.
 $\gamma_{num}$ and $\gamma_{exp}$ denote our numerical results and experimental results by Childers and Tough \cite{tough,childers}, respectively.\label{gamma}}
\end{table}
\end{center}

\begin{figure}[h]
\begin{center}
\scalebox{0.44}{\includegraphics{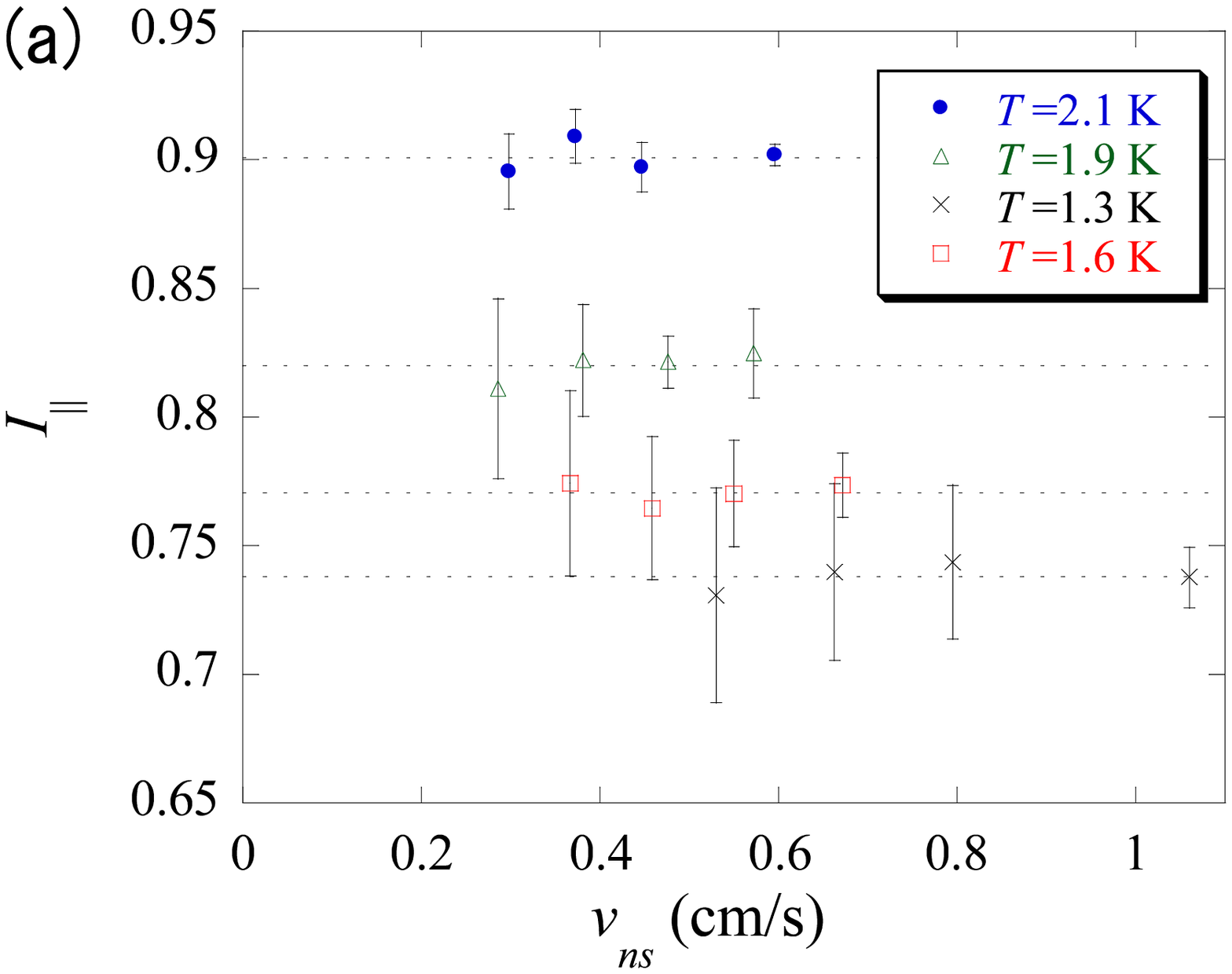}}
\scalebox{0.44}{\includegraphics{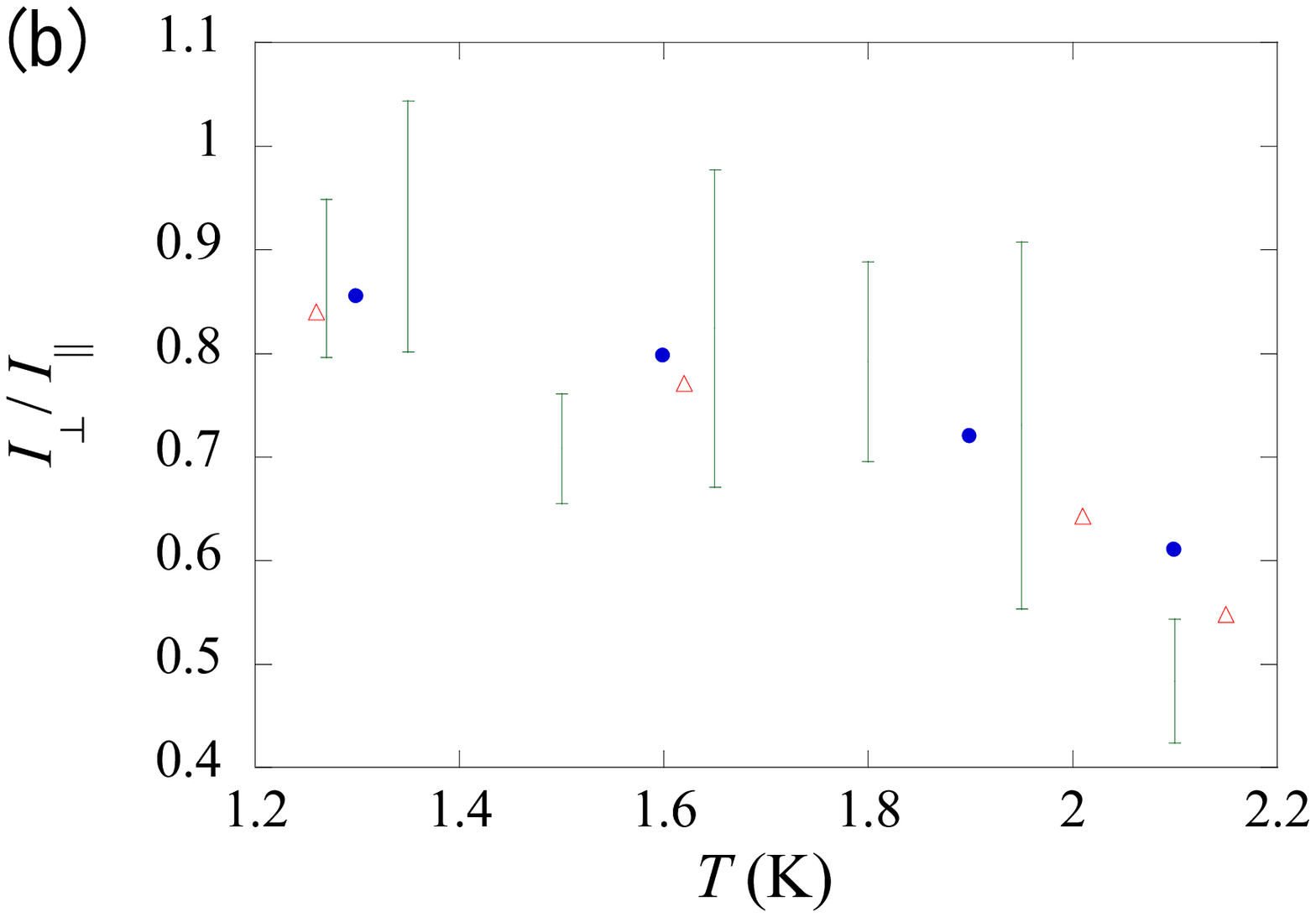}}
\caption{(a) Anisotropy parameter $I_\|$ as a function of $v_{ns}$. The error bars represent the standard deviation. (b) Anisotropy ratio $I_{\bot}/I_{\|}$ as a function of temperature obtained by us (dots), Schwarz\cite{schwarz88} (triangles) and experimentally\cite{anisotropy} (vertical bars). 
\label{ani_const}}
\end{center}
\end{figure}

 Figure \ref{ani_const}(a) shows the anisotropy as a function of $v_{ns}$ and $T$. The anisotropy is almost independent of $v_{ns}$ and is dependent on $T$, in agreement with experimental observations \cite{anisotropy}. The anisotropy ratio $I_{\bot} / I_{\|}$ has been measured experimentally \cite{anisotropy} and estimated by numerical simulation \cite{schwarz88}. An isotropic vortex tangle yields $I_{\bot} / I_{\|}=1$. If the vortex tangle consists entirely of curves lying in planes normal to ${\bf v}_{ns}$, then $I_{\bot} / I_{\|}=1/2$. We show this anisotropy ratio in Fig. \ref{ani_const}(b). The anisotropy increases with increasing temperature because the mutual friction increases. The steady state shows slightly higher values of $I_{\bot} / I_{\|}$ than those obtained by Schwarz. This is why the vortex interaction reduces the anisotropy, as discussed in Sec. I\hspace{-.1em}V. 

All our numerical results are in reasonable agreement with experimental results. This means that our model realistically simulates counterflow turbulence. The steady state was obtained without using an artificial mixing procedure.
\section{Validity of the LIA}
\begin{figure}[h]
\begin{center}
\scalebox{0.4}{\includegraphics{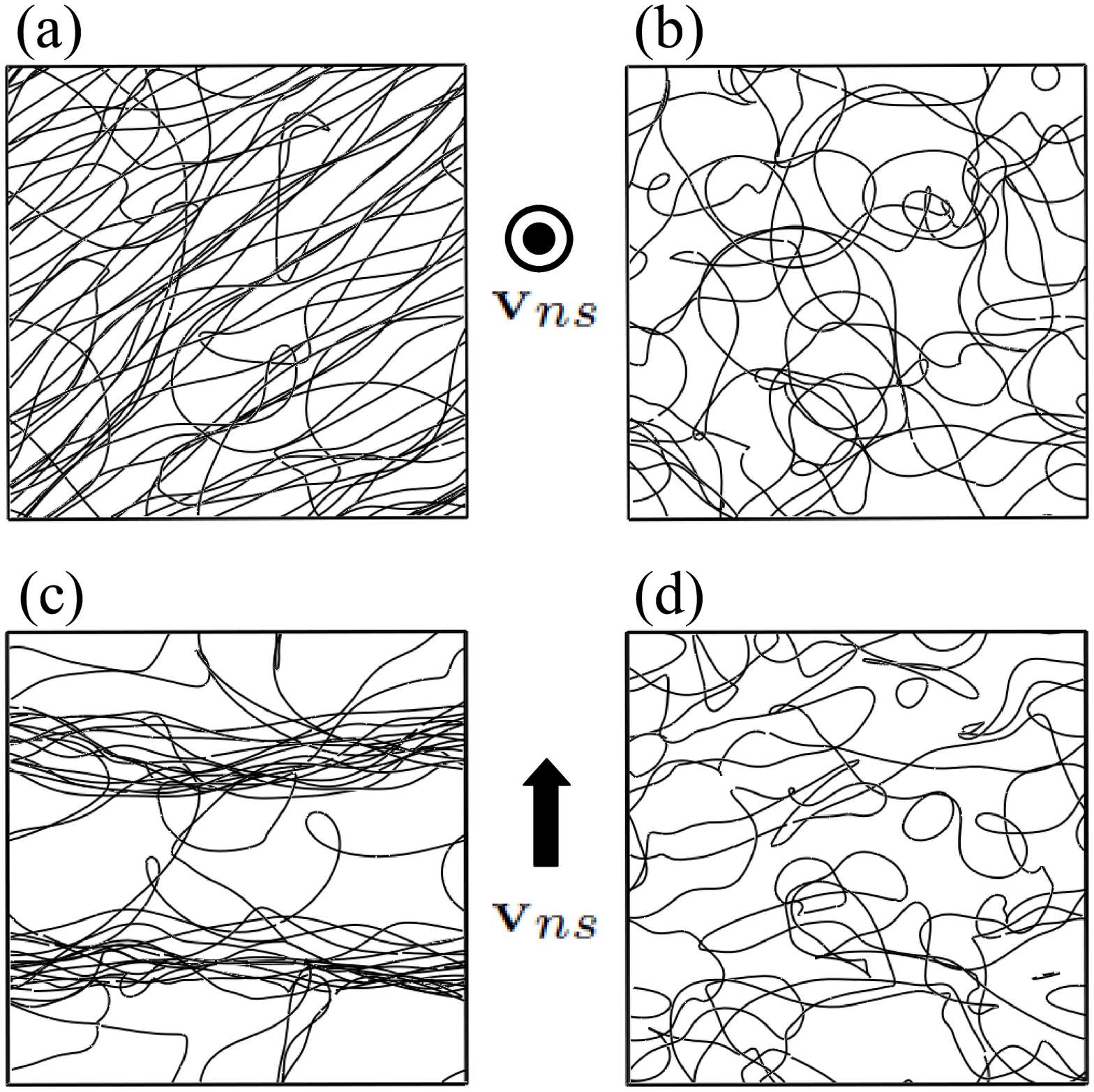}}
\caption{(a) Side and (c) top views of the LIA calculation. (b) side and (d) top views of the full Biot--Savart calculation. All figures are at $t=18.6 \,{\rm s}$. The system is a $(0.2cm)^3$ cube. Applied normal fluid velocity is $v_{ns}=0.55\,{\rm cm/s}$.\label{T16}} 
\end{center}
\end{figure}

\begin{figure}[h]
\begin{center}
\scalebox{0.45}{\includegraphics{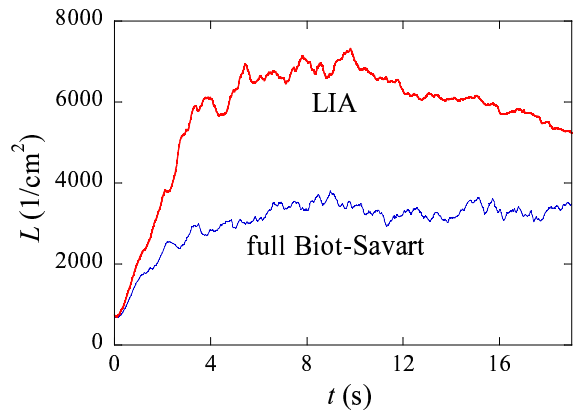}}
\caption{(color online). Comparison of the time evolution of the vortex line density $L(t)$ for the simulation of Fig. \ref{T16}.\label{T16linedensity}} 
\end{center}
\end{figure}

\begin{figure}[h]
\begin{center}
\scalebox{0.45}{\includegraphics{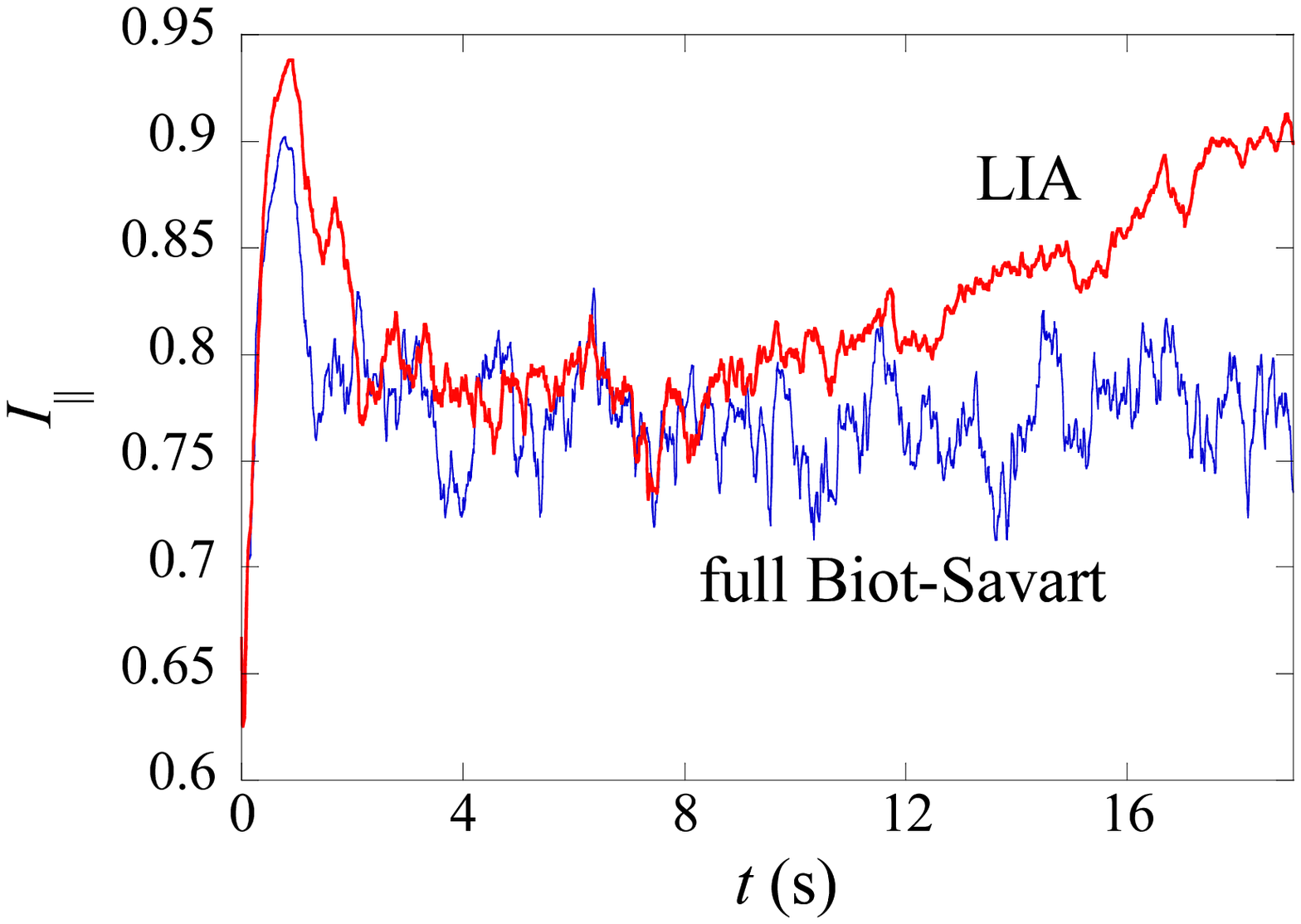}}
\caption{(color online). Comparison of the time evolution of the anisotropy parameter $I_{\|}$(t) for the simulation of Fig. \ref{T16}.\label{T16anisotropy}} 
\end{center}
\end{figure}

Previous studies of counterflow turbulence with the LIA have encountered some serious difficulties. Schwarz could not obtain the statistically steady state under periodic boundary conditions without using a mixing procedure, as discussed in Sec. I.
 Kondaurova et al. could obtain the steady state using a different reconnection procedure from us, which is discussed in Sec. V.
However, their values of $\gamma$ were three times larger than experimentally measured values. These difficulties probably arise from the LIA.
In this section, we compare two calculations, namely the LIA (Figs. \ref{T16}(a),(c)) and the full Biot--Savart law (Figs. \ref{T16}(b),(d)).
 We run both calculations at the same condition, $T=1.6\,{\rm K}$ and $v_{ns}=0.367 \,{\rm cm/s}$. The time evolution of $L(t)$ and $I_{\|}(t)$ is shown in Figs. \ref{T16linedensity} and \ref {T16anisotropy}. 

\begin{figure}[h]
\begin{center}
\scalebox{0.2}{\includegraphics{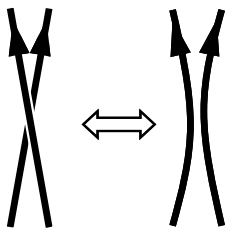}}
\caption{Reconnection of almost parallel vortices. \label{parallel}} 
\end{center}
\end{figure}

The dynamics of vortices by the LIA is qualitatively similar to that with the full Biot--Savart law in the first and second stages described in Sec. I\hspace{-.1em}I\hspace{-.1em}I.
The vortex tangle appears to enter the steady state for $4 \leq t \leq 8 \,{\rm s}$. However, the VLD is about twice that for the full Biot--Savart law, as shown in Fig. \ref{T16linedensity}.
This difference is due to the absence of interaction between vortices; the interaction in general works strongly immediately before and after reconnections. In the case of the full Biot--Savart calculation, through the mutual friction the interaction between vortices tends to separate two parallel vortices and bring two antiparallel vortices closer to reconnection. Reconnection generally creates sharp cusps\cite{schwarz85} in the vortex lines, so that the vortices separate rapidly with a large self-induced velocity.
However, the LIA calculations result in very different reconnection dynamics. Reconnection between parallel vortices occurs frequently in the LIA, whereas they occur very little in the full Biot--Savart calculation due to the interaction.
After reconnection between parallel vortices specific to the LIA calculation, the newly created vortices cannot separate rapidly because the radius of curvature is large, as shown schematically in Fig. \ref{parallel}, which tends to get the vortices together. 
 Also, the resulting vortices are affected little by the decay mechanism described in Sec. I\hspace{-.1em}I\hspace{-.1em}I.
Consequently, the average distance of the LIA is smaller and the VLD is larger than those of the full Biot--Savart calculation.
 
\begin{figure}[h]
\begin{center}
\scalebox{0.4}{\includegraphics{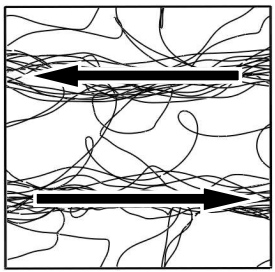}}
\caption{Vorticity direction of bundles in the layer vortices shown in Figs. \ref{T16}(a) and (c). \label{direction}} 
\end{center}
\end{figure}

The expansion by mutual friction tends to straighten vortices perpendicularly to ${\bf v}_{ns}$, but reconnection suppresses this tendency by creating a low radius of curvature.
However, as discussed above, in the LIA calculations, parallel vortices tend to come together due to the absence of interaction; hence, vortices cannot create a small radius of curvature, which gradually straighten and align vortices. Vortices eventually commence forming a bundle structure composed of parallel straight vortices, as shown in Figs. \ref{T16}(a) and (c), which is bunches of straight vortices, not turbulence. In this state, the vorticity directions of the vortex bundles differ from one layer to another, as shown in Fig. \ref{direction}. 

The LIA calculation gives very different vortex properties from the full Biot--Savart calculation. We can thus conclude that the LIA is unsuitable for simulating counterflow turbulence.
The interaction between vortices plays an important role in making turbulence uniform. It probably applies to any kind of quantum turbulence.

\section{Conclusions}
In this study, we investigated thermal counterflow turbulence using the vortex filament model. The full Biot--Savart law was used in our calculation, unlike previous studies by Schwarz \cite{schwarz88}. We obtain the statistically steady state without using the mixing procedure that Schwarz used to sustain the steady state. Our numerical results reveal the characteristic relation $L=\gamma^2 v_{ns}^2$, which has been observed in many experiments, and the parameter $\gamma$ agrees with experimentally measured values. Also, the anisotropy parameters are in reasonable agreement with experimental observations. 

To investigate the validity of the LIA, we compared the LIA with the full Biot--Savart calculation. Kondaurova et al. mention that the reason why Schwarz encountered difficulties is due to the reconnection procedure; they obtained the steady state using the LIA and a different reconnection procedure from us and Schwarz \cite{kondaurova}. They assume that reconnections occur only when the vortices are expected to cross each other. Probably the details of the reconnection procedure are not relevant to statistical quantities such as the VLD and the anisotropy parameters.
 Kondaurova et al. encountered another difficulty in that the value of the parameter $\gamma$ is very large. Our comparison between the LIA and the full Biot--Savart reveal that the difficulties encountered by Schwarz and Kondaurova et al. originate from the LIA.
The steady state of turbulence cannot be sustained in LIA calculations because most vortices gradually straighten and lie in the planes normal to $v_{ns}$. Since the full Biot--Savart calculation includes the interaction between vortices, our findings imply that the interaction between vortices is essential for turbulence simulations.
 
Paoletti et al. succeeded in visualizing counterflow using solid hydrogen particles\cite{paoletti}, and they obtained a bimodal distribution for the particle velocity. Paoletti et al. expect that particles dragged by the normal fluid and particles trapped in the vortex tangle contribute to this bimodality, but this still remains to be confirmed.
 Further study is required to understand this observation. 

\section{Acknowledgement}
 S. F. acknowledges the support of a Research Fellowship of the Japan Society for the Promotion of Science for Young Scientists (Grant No. 217762). M. T. acknowledges the support of a Grant-in Aid for Scientific Research from JSPS (Grant No. 21340104) and a Grant-in-Aid for Scientific Research on Priority Areas from MEXT (Grant No. 17071008).

\end{document}